# Plasma absorption levelling by phase coherence in meta-surface stacked structure

Xing-Yu Xu,[1,2] Zhong-Zhu Liang,[3,] Li Qin[2] and Xue-Mei Su[1,*]

[1] *Physics College, Jilin University, Changchun 130012, P. R. China*
[2] *State Key Laboratory of Applied Optics, Changchun Institute of Optics, Fine Mechanics and Physics, Chinese Academy of Sciences, Changchun, Jilin, 130033, China*
[3] *College of Physics, Northeast Normal University of China, Changchun 130024, P. R. China*

*suxm@jlu.edu.cn

**Abstract**: In this paper, we propose a MIM (metallic metasurface-insulator-metal) stacked structure to realize perfect absorption in mid- and far infrared bandwidth. A large number of metallic composite metallic units placed on a uniform layer of insulator Ge which is deposited on a uniform metallic Ti surface. Each of units consists of four right-angled triangular cubes in middle of four sides of a square. Three resonant absorption peaks can be discrete or levelling dependent to thickness of cubes. Once metallic cubes are thinner than its skin depth, an effective ultra-broadband absorber is realizable with average absorption over 90% percent ranging from 8-14 μm. We build a four-level cavity-dipole interacting model to explain phenomena of plasma absorption levelling. This is created by strong phase coherence from meta-surface since three hybrid modes are coupled by common cavity photons in arrays of micro-cavities between top and below metallic layers. The structure is polarization-selective to infrared light. It has great potential in infrared detection and imaging.

Pacs number: 42.50.Gy, 42.50.Pq, 52.38.-r

The authors [1] stated that an incident light can experience diffraction through a single metallic slit at subwavelength and drive excitation of surface plasmon polaritons (SPP) inside the slit. The SPP modes are resulting from free charge plasmon oscillation at metal-dielectric interface. The excitation efficiency is proportional to diffraction factor of the single slit, sinc ($\pi a\, \theta / \lambda$), as in formula (6) [1] where a: slit width, $\theta$ : diffractive angle. It is more than a traditional diffraction of single slit due to excitation of bound SPP modes. The metallic slit can be made by shallow corrugation in lamellar structure [2] which easily conducting incident light. It determines the wavelengths of SPP waves by shape, size, orientation, dielectric constant [3] and refractive index of dielectric medium below and inside it. The real part of dielectric constant $\varepsilon$ [4] of metal Ti is larger than refractive index of Ge $n_2$ [5] and therefore SPP waves is decaying rapidly in metal and bound to the interface.

A fundamental mode (TE mode) in a single slit is derived in analytical forms (Ez, Hy is not zero) [1, 6] with a largest wavelength and thickness-independent. Another thickness-dependent [7-9] resonant mode (Ex, Hy is not zero) is not involved in [1] whose magnetic component Hy can be expanded into several Fabry-Perot (F-P) modes [10]. The resonant wavelength is thickness-dependent with a difference from Fabry-Perot modes relative to the value "w/t" (t is thickness) [7]. That is a TM mode of electromagnetic field from SPP excitation in thickness direction with lower efficiency. However, it can be enhanced greatly if a meta-surface [11, 12] is utilized to create quantum interference between SPP raditions from majorities of single slits placed periodically. The basic interference effect has been proved in experiment of metallic double-slits [13].

A meta-surface MIM structure usually consists of tri-layers [14, 15], a top layer of nano- or micro-metallic particles or holes, a bottom uniform metallic ground plane separated by a layer of uniform dielectric. It can compensate mismatching of moment [16] between SPP and incident light based on effect of surface lattice comparing with those of few metallic holes on top surface [17]. Recently, MIM structure is widely used to make perfect absorbers in solar cell [18] and imaging equipment [19] for easily coupling between SPP wave and free propagating light in atmosphere [20, 21]. The magnetic/electric field can be strongly localized in the spacing dielectric layer [20]. Most of MIM absorbers exhibit higher absorption based on interference from total lattice sum but in a comparably narrow absorption band [21]. To conquer this limitation, some composite units are operated at several wavelengths [23, 24]. With proper parameters, absorption levelling [25] can be realized in a comparatively wide range.

In this paper, a composite unit is devised by four triangular cubes as a unit to build a 2D meta-surface in MIM stacked structure to realize perfect far-infrared absorber in ultra-broad bandwidth. This type of absorber is advantageous to image contrast enhancements in precise target detection [26] and facial recognition [27] in natural environment. The meta-surface with arrays of units can lead to superior performances for each of which places four asymmetric triangles in a symmetric way. The incident light propagating in vertical direction can be diffracted into the insulator layer with assistance of meta-surface. Quantum coherence takes influence on plasma absorption [28] in meta-surface structure. Here we find that in some conditions, MIM structure is functioned as a waveguide cavity to couple cavity photons with dipoles at hybrid modes and induce plasma absorption levelling (PAL).

The schematic of the proposed absorber is shown as in Fig. 1 (a) which is formed by metallic meta-surface of triangular Ti cubes, a layer of Ge dielectrics and a layer of uniform Ti metal. On the right is the enlarged diagram of a single unit cell. It includes four identical isosceles right-angled cubes placed in a symmetric manner as the area in yellow line. They are located in sides of the square units and apexes of every two triangles are opposite. Its hypotenuse length d is in order of μm but smaller than wavelength of input light. The third subgraph in Fig. 1(a) is lateral diagram of the absorber where $t_1$, $t_2$ and $t_3$ correspond to the thicknesses of the cube, dielectrics and metal film on the bottom, respectively. The thickness $t_3$ is large enough surrounding by the air to prevent light from leaking. The absorption spectrum of the absorber at $t_1$=30 nm and 200nm are depicted in Fig.1 (b) and 1 (c), respectively with the other parameters identical. The perfect absorber possesses over 90% absorptivity levelling in range of 8~14 μm for thinner metallic cubes ($t_1$=30 nm) as shown in Fig 1(b).

For thicker cubes ($t_1$= 200nm), it exhibits still three resonant peaks but comparatively low absorption as in Fig. 1(c). Resonant wavelengths are dependent to the effective lengths of the metallic slots, as indicating as the orange dotted quadrilateral in Fig. 2(a)-(c). The corresponding $\lambda_1$, $\lambda_2$ and $\lambda_3$ are equal to twice of the effective lengths $\Lambda_{1,2,3} \sim d/\sqrt{2}, d, \sqrt{2}d$, multiplied by refractive index at interface [1] ($\varepsilon n_2^2/(\varepsilon + n_2^2))^{1/2}$; d is the hypotenuse length of the metallic cubes. From view of periodical arrangement in Fig. 2(a)-2(c), meta-surface consists of two types of dipoles or "atoms", those are quasi-diamonds (red box) formed by two identical triangles with X- and Y- orientation splits, respectively. The SPP excitation at resonant modes $\lambda_1$, $\lambda_2$ are mainly resulting from polarization of Y-split atoms, while mode at $\lambda_3$ is from excitation of X-split atoms. This creates SPP waves propagating or localized in x-z and y-z planes in space between metallic cubes and bottom film as in Fig. 2(d)-(f) by the same parameters as Fig. 1(c).

Here polarization direction of the incoming light is chosen as x axis. It can pass through MIM structure benefiting from traditional diffraction at subwavelength metallic slots and then propagate in Ge layer at a small diffraction angles to the bottom metal and finally reflecting to impinge the metal cubes from below. According to principles of gratings diffraction, the half-angle-width of zero-order diffraction θ ~ λ/(N*u) is small enough since the number of units N is so large (u is periods of metallic gratings). The optimum depth of Ge layer decides by effective optical length of incident light to satisfy $2n_2t_2 \approx \lambda_2/2$ in order for a maximum diffraction intensity to drive SPP excitation at $\lambda_2$ directly by incident light. The other two modes can be excited in nonlinear processes [7]. According to absorptions spectra in curves (a)-(d) in Fig. 3(a) by FDTD software, the resonant modes can be thickness-dependent ($\lambda_1$, $\lambda_2$) or -independent ($\lambda_3$) exhibiting characteristics of TM or TE modes at discrete wavelength [17, 29].

It was assumed that every unit are drive by incident light identically, the total polarization of any unit $P= -AE_0/((\omega - \omega_{spp} - \Omega_i) + i(\gamma_{spp} + \gamma_i))$ is dependent to susceptibility of itself by Drude model $\alpha_s = -A/((\omega - \omega_{spp}) + i\gamma_{spp})$; (A is SPP oscillation strength [17]) and impact from other units on meta-surface, namely lattice sum, S=$\sum S_j$; (j=1,2,3…N-1). $S_j$ is dependent to distance $r_j$ between the $i^{th}$ and $j^{th}$ units [29]. The impact embodied in frequency shift $\Omega_i = ARe(S)$ and increment of damping rate $\gamma_i = AIm(S)$. When $r_j \gg \lambda_{SPP}$, $S_j$=0 indicating none coupling between $i^{th}$ and $j^{th}$ units. While $r_j < \lambda_{SPP}$, $S_j$ is a limited complex number dependent to magnitude $r_j$, as well as radiative angle between $r_j$ and dipole moment. S is fast convergent since only a few of none-zero and limited terms to be summarized. However, this is sensitive to relative phases dependent to atomic distances. That is basic elements in meta-surface for dipole-dipole interaction [30]. When distance of neighboring unit "w" is around half of $\lambda_1$ as curves (c) and (d) in Fig. 3(b), phase interference is created in absorption cancellation or enhancement at $\lambda_1$ as in atomic systems [31] even though collective damping rate $\gamma_i$ is as large as in dozens of THz.

Once thickness of metallic cubes less than skin depth ($t_{skin}$) of metal Ti, electrons passing through the metallic cubes, positive and negative charges can be populated in upper and lower surfaces of the cubes. Another SPP resonant modes is excited within air-metal slit in lateral plane with effective lengths $\Lambda_4$=w*$n_{in2}$ (w, width of a unit cell, $n_{in2}$=($\varepsilon n_0^2/(\varepsilon + n_0^2))^{1/2}$). This leads to three hybrid modes whose effective lengths $\Lambda_{1h,2h,3h} = \sqrt{\Lambda_{1,2,3}^2 + \Lambda_4^2}$. The localized fields are distributed as shown in Fig.4. The hybrid modes exhibit characteristic of superposition of TE/TM or TM/TM modes and differently polarized in x and y directions. Under condition that $t_1 < t_{skin}$, high absorption and absorption levelling are achieved as curves (e) and (f) in Fig. 3.

We start to build field amplitudes equations in a four-level system describing states of dipoles as in Fig. 5(a) to indicate physical essence of cavity induced plasma absorption levelling (PAL). Based on a

MIM structure in Fig. 1(a), metasurface includes a lot of identical units in a 2D array of microcavities in thickness direction between top and lower metallic layers to support coupling between cavity photon and SPP waves at resonant modes. Each microcavity is thin enough in its fundamental mode and populated cavity photons with frequency ω$_k$=($\omega_c$ − 2Jcosk) [32] in a bandwidth 4J around center frequency $\omega_c$ by considering impact of others in periodical arrangement (J is coupling strength between two microcavities at different positions). The photons are described by annihilation operator **a** (creative **a**$^\dagger$) and average field amplitude a(ω) [33]. The dipoles or atoms at excitation modes are described by annihilation operators **b**$_j$ (**b**$_j^\dagger$) (j=1, 2, 3) in states |j> at resonant frequency ω$_j$=2 π c/ λ $_j$ of transition from unexcited state |0>. The average field amplitudes b$_j$(ω) is a variable in frequency space with respect to Fourior variable ω. The dipoles at three excited states interact with the intracavity photons at any position by coupling parameters $\kappa_i$. The field amplitudes can be described by Langevine equations in frequency space [28, 33]

$$\begin{bmatrix} a \\ b_1 \\ b_2 \\ b_3 \end{bmatrix} = - \begin{bmatrix} \delta + i\gamma_c & \kappa_1 & \kappa_2 & \kappa_3 \\ \kappa_1 & (\delta - \Delta_{c1}) + i\gamma_1 & 0 & 0 \\ \kappa_2 & 0 & \delta + \delta_c + i\gamma_2 & 0 \\ \kappa_3 & 0 & 0 & (\delta + \Delta_{c2}) + i\gamma_3 \end{bmatrix}^{-1} \begin{bmatrix} 0 \\ 0 \\ gE_0 \\ 0 \end{bmatrix} \quad (1)$$

$$b_2 = \frac{gE_0(\kappa_3^2(\delta-\Delta_{c1}+i\gamma_1)+\kappa_1^2(\delta+\Delta_{c2}+i\gamma_3)-A(\delta+\Delta_{c2}+i\gamma_3))}{B(\delta-\Delta_{c1}+i\gamma_1)-(A-\kappa_1^2)(\delta+\delta_c+i\gamma_2)(\delta+\Delta_{c2}+i\gamma_3)} \quad (2)$$

where A = $(\delta - \Delta_{c1} + i\gamma_1)(\delta + i\gamma_c)$; B = $-\kappa_3^2(\delta + \delta_c + i\gamma_2) - \kappa_2^2(\delta + \Delta_{c2} + i\gamma_3)$; variable $\delta$ stands for frequency detunings between incident light and cavity photon; $gE_0$ is Rabi frequency for describing interaction between dipoles and incident field, whose amplitude is $E_0$. $\delta_c = (\omega_2 - \omega_c)$ is its detuning from the bright hybrid mode, $\Delta_{c1,2} = \Delta_{1,2} \mp \delta_c$ and $\Delta_1 = (\omega_2 - \omega_1)$, $\Delta_2 = (\omega_3 - \omega_2)$ are large enough in the four level cavity-dipole system in a similar way with three subsystems for gain levelling in Er$^{3+}$ doped fibers [34]. There one of subsystems is resonant to the driving field and the other two un-resonant with large detuning $\pm\Delta$.

In order for parameters in equations (1) and (2), we describe the system in a model of waveguide cavity as in Ref. [34]. Under J>> g$_0$, the coupling strength of atom-cavity is decided by a phase-dependent term as $\left|\frac{g_{0i}^2}{v(\delta)}\right| e^{i(k_0 r_j)}$, which is involved in individual atom-cavity g$_{0i}$=| μ $_i$|$\sqrt{\frac{\omega_c}{2\hbar\varepsilon_0 V}}$ (i=1,2,3) [35] and as well as impacts of other atoms away from distance r$_j$. The collective coupling strength is dependent to spontaneous emission coherence between "atoms" in waveguide cavities [36], including amplitude effect with group velocity $v(\delta) \approx \sqrt{(2J)^2 - (\delta_c)^2}$ and phase term $e^{i(k_0 r_j)}$. The phase shift $k_0 r_j = k\beta_0 r_j/2$ [37] is caused by incident field to control intracavity EIT medium pulling frequencies of photons and corresponding atomic transitions into a broad bandwidth. The wavevector of incident light k = $2\pi/\lambda_0$, $\beta_0$ is corresponding to real part of atomic susceptibility, here according to dispersive relation in arrays of atom-cavity system $\beta_0 = (1 - \arccos\left(\frac{\Delta_c}{2J}\right)/\pi)$. The coupling strength [30, 38] in arrays of cavity-dipole system κ$_i$ =$\sum_{j=1}^{\pi/k_0 r_{12}} S_j$ and $S_j = \frac{3\gamma_i}{2}\left(-\frac{cos\xi}{\xi} + \frac{sin\xi}{\xi^2} + \frac{cos\xi}{\xi^3}\right)$; $\xi = k_0 r_j$ (angle of dipole moment μ and vector r$_j$, ɑ = π /2). We use resonant atoms, $\Delta_c = 0$, to prove the coupling strength $\kappa_i$ and find it is accordance with results using a model of a single atom interacting with a mirror to describe the system [38]. The damping rates of collective atomic spontaneous emission are estimated γ $_1$=0.85 γ , γ $_2$=1.25 γ , γ $_3$=1.15 γ and γ =10 THz, from Fig. 1(c). γ $_c$ is in the same order with atomic transitions ~1.2 γ and J=25 γ , g$_{0i}$= γ $_i$.

The population probability |b$_2$|$^2$ and absorption spectra are depicted in Fig.5 (b)-(d) to show effect of phase coherence on high absorption and modes combination under different conditions of frequency detunings and coupling strengths. The three hybrid modes interact with cavity photons in different way due to phase coherence of a number of metallic units in arrays of microcavity. The absorption rates can be increased even coupling with atoms in large detunings for collective interferent effect in the coupling constants. The incident light exerts impact to the system as that of the control laser in Ref. [39], can lead to combination of the hybrid modes and realization of high absorption levelling. It takes effects on resonant ($\Delta_c = 0$) and ($\Delta_c \neq 0$) un-resonant atoms differently on frequency detunning [39], and so the optimum frequency detunings for mode combination $\delta_c$ is in middle of the object modes. Curves(a)-(d) in Fig. 5(b) are drawn with different frequency detunings $\delta_c$ in conditions $\Delta_1 = 4.1 \gamma$, $\Delta_2 = 4.2 \gamma$, strong dipole-dipole interaction at distance w= 2.25 μm. It shows influence from frequency detuning on phase coherence, modes combination and absorption levelling using other parameters are the same as in Fig.1 (b). Curve (a) exhibits combinations of three modes with each other caused by a common cavity

photon driven by a nearly resonant incident light with a small detuning $\delta_c = 0.49\gamma < \frac{\gamma c}{2}$. The absorption leveling is similar as gain levelling in Er$^{3+}$ doped fibers [33] but without populations transfer in absent of incoherent pumping. In curves (b) and (c) in conditions $\delta_c = 0.66\gamma$ and $0.82\gamma$, the second mode keeps coupling with the first one by "bright" cavity photon illuminated by incident light. The absorption levelling still happened but the ranges is decreased comparing with smaller detuning. Until $\delta_c = 0.98\gamma > \gamma_1$ as for curve (d), which is the frequency detuning of incident light beyond extent of modes combination [39] and result in three discrete absorption peaks. However, absorption rates is high for cavity induced coupling with each atomic transitions. Fig. 5(b) indicates absorption leveling in a wider range requires combination of two or three hybrid modes by phase coherence. Correspondingly in Fig. 5(c), absorption spectra at several insulator thickness t$_2$ are accordant with numerical simulation in Fig. 5(b) by using software of FDTD solution. It shows that levelling bandwidth decreased with frequency detuning being enlarged even by strong phase coherence at distance w= 2.4 μ m.

Plasma absorption levelling vs. different unit widths is shown as curves (e)-(g) in Fig.5(b) and Fig. 5(d) to show influence of both coupling strength and frequency detunings. By keeping parameters of metallic cubes but increasing distance of adjacent unit "w", dipole-dipole effect is weakened in fast way by $\sim (w/\lambda_h)^{-3}$ and on the other hand, frequency detunings being enlarged between hybrid modes and cavity photon. Shown as in Fig.5(b) and Fig. 5(d), smaller distance of the neighboring units leads to stronger phase coherence of cavity photon, absorption is levelling in a ultrawide range. With a larger distance, such as w=2.8 μm, there are two TM hybrid modes being coupled by cavity photon, and so absorption levelling is in a smaller range. It finally disappeared leading to three discrete peaks with much larger unit widths.

In addition, we find that levelling range is independence to the polarization angle of incident light due to symmetric arrangements of every meta-surface unit (not shown). This make it easily utilized in atmospheric environment whether oblique incidence or not.

In a word, we propose and prove a metallic metasurface-insulator-metal (MIM) stacked structure to achieve an ultra-broadband levelling absorber in long wavelength infrared regime. The absorption peaks at resonance are discrete or levelling dependent on thickness of the metallic cubes on top surface. We prove phase coherence in waveguide cavity and its applications in high absorption and absorption levelling. This is resulted from dipole-dipole interaction between three hybrid modes and bright arrays of cavity between top and below metallic layers. The levelling ranges can be controlled by tuning each triangle unit size, distance, thickness and that of insulator Ge. This MIM structure is polarization selective by polarized directions of incident light.

Figure caption

Fig. 1: (color online) (a) Schematics of the proposed broadband meta-material absorber. The yellow arrow pointed subfigure is a magnified unit cell; below is a lateral diagram showing thicknesses of metal cube, Ge and bottom Ti film as $t_1$, $t_2$ and $t_3$. (b) and (c) absorption spectrum of proposed absorber with $t_1$ = 30nm, 200nm, respectively; other parameters are w = 2.4μm, d= 1.2μm, g=1/√2μm, $t_2$ = 530nm, $t_3$ = 200nm.

Fig. 2: (color online) (a)-(c) four periods of unit cell of the meta-surface to show diamond-like dipoles. The dotted quadrilaterals indicate three types of metallic slots based on composite meta-surface. (d)-(f) magnetic field distributions in x- z and y- z planes in one unit at $\lambda_1$, $\lambda_2$, $\lambda_3$= 6.83μm, 9.05 μm, 12.64 μm by parameters in Fig. 1(c).

Fig. 3: (color online) (a) absorptions of the proposed absorber vs. thicknesses of the metallic cubes t1 and w = 2.4 μm. (b) absorption vs. width w for thick cubes, $t_1$ =200nm. other parameters are d= 1.2μm, g=1/√2μm, $t_2$ = 530nm, $t_3$ = 200nm.

Fig. 4. (color online) (a-c) Electric and (d-f) magnetic field distributions for thin cubes in x- z and y- z planes, respectively by the same parameters as in Fig. 1(b).

Fig. 5: (color online) (a) energy level diagram for a simplified four-level system. (b) calculated results of population probability $|b_2|^2$ versus frequency detuning δ to demonstrate plasma absorption levelling; curves (a)-(d) is depicted by modulating frequency detunings between cavity photon and dipoles under strong dipole-dipole interaction at unit distance w=2.25 μm ; curves (e)-(g) is for modulation of distance of neighboring units, "w". (c) and (d) numerical results of absorptions spectra by different thickness $t_2$ and distances "w", respectively by FDTD solution software.

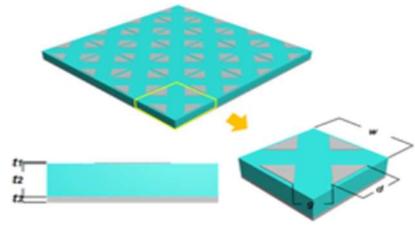

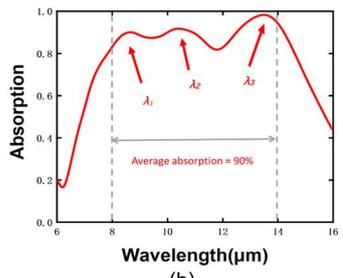
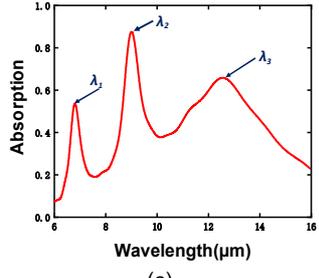

(a)

(b)   (c)

(a) (b) (c)

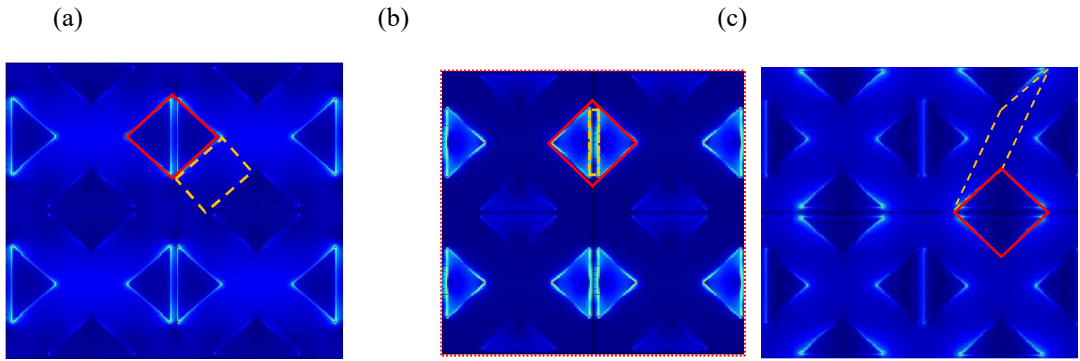

(d) (e) (f)

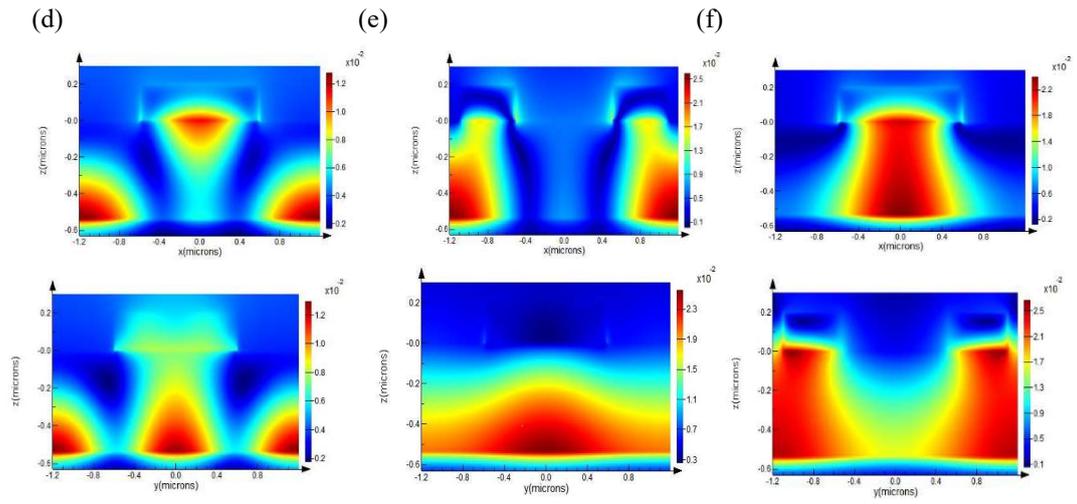

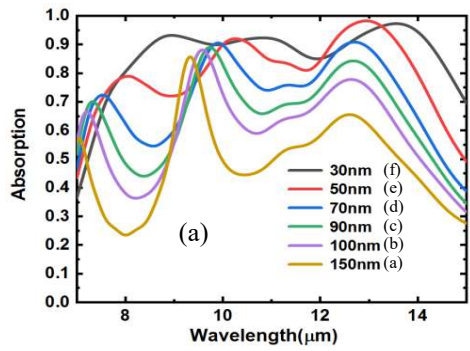

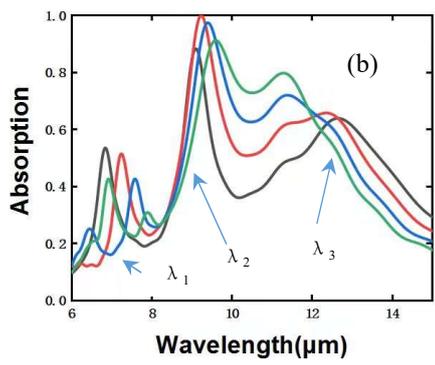

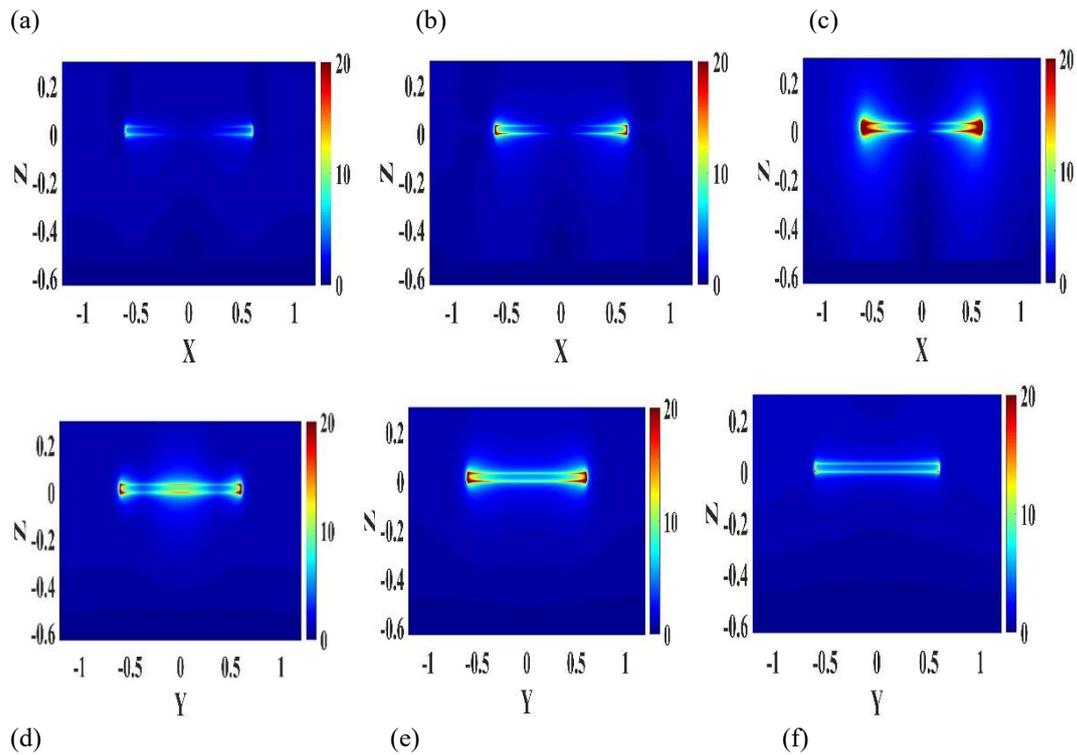

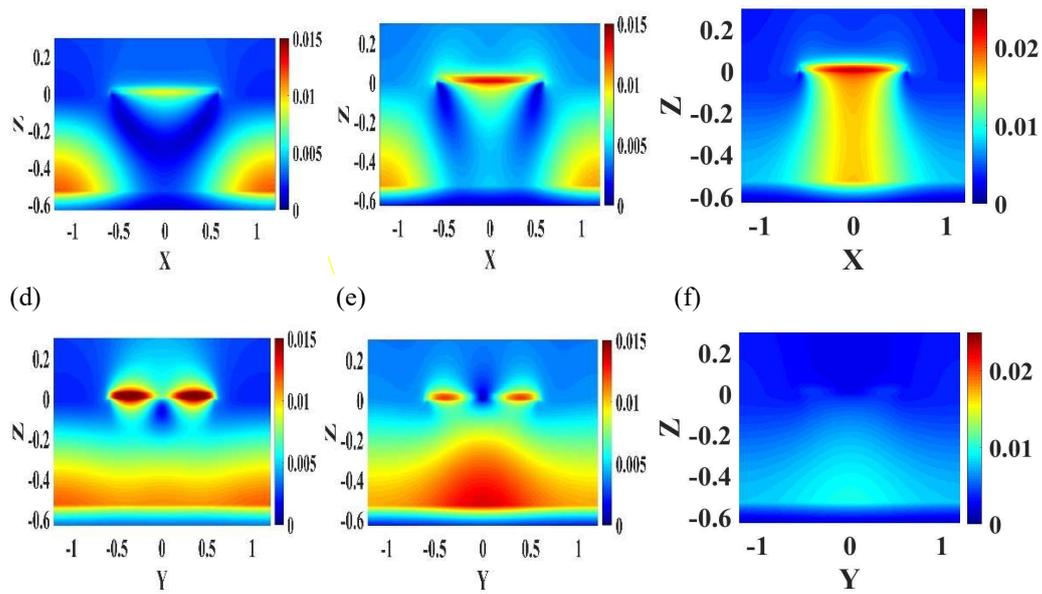

(a)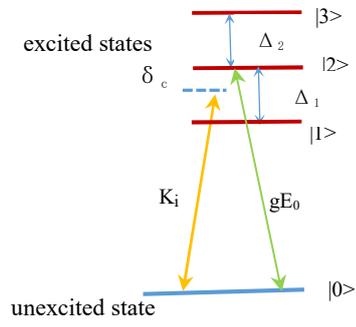 (b)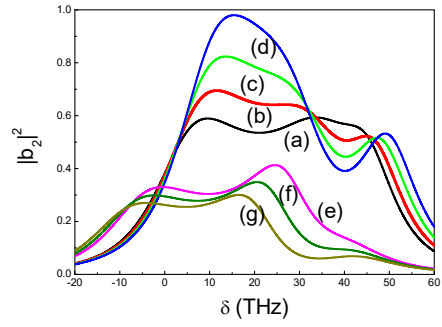

(c)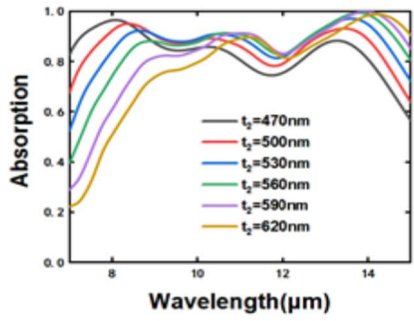 (d)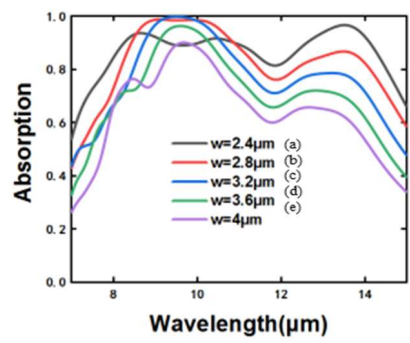